\documentclass[twocolumn,showpacs,preprintnumbers,nofootinbib,amssymb]{revtex4-1}

\usepackage[dvipdfm]{graphicx} 
\usepackage{epsfig, amssymb} 
\usepackage{amsmath, amsfonts}
\usepackage{bm} 

\usepackage[linktocpage]{hyperref}
\usepackage[caption=false]{subfig}
\usepackage[usenames]{color}

\def\beq{\begin{equation}}
\def\eeq{\end{equation}}
\def\beqa{\begin{eqnarray}}
\def\eeqa{\end{eqnarray}}

\begin{document}

\title{\large Black holes in anti-de Sitter: quasinormal modes, tails and tales of flat spacetime}

\author{Vitor Cardoso} 
\email{vitor.cardoso@ist.utl.pt}
\affiliation{CENTRA, Departamento de F\'{\i}sica, Instituto Superior T\'ecnico, Universidade de Lisboa,
Avenida Rovisco Pais 1, 1049 Lisboa, Portugal.}
\affiliation{Perimeter Institute for Theoretical Physics Waterloo, Ontario N2J 2W9, Canada.}
\affiliation{Department of Physics and Astronomy, The University of Mississippi, University, MS 38677, USA.}

\author{Gaurav Khanna}
\email{gkhanna@umassd.edu}
\affiliation{Physics Department, University of Massachusetts Dartmouth, North Dartmouth, MA 02747, USA.}


\begin{abstract} 
Black holes (BHs) in asymptotically anti-de Sitter (AdS) spacetimes have been the subject of intense scrutiny, including detailed
frequency-domain analysis and full nonlinear evolutions. Remarkably, studies of linearized perturbations in the time-domain
are scarce or non-existing. We close this gap by evolving linearized scalar wavepackets in the background of rotating BHs
in AdS spacetimes. Our results show a number of interesting features. Small BHs in AdS behave as {\it asymptotically flat BHs} for early/intermediate times, displaying the same ringdown modes and power-law tails. As the field bounces back and forth between the
horizon and the timelike boundary it ``thermalizes'' and the modes of AdS settle in. Finally, we have indications that wavepackets in the vicinity of fastly spinning BHs grow {\it exponentially} in time, signalling a superradiant instability of the geometry previously reported through a frequency-domain analysis.
\end{abstract}

\maketitle

\section{Introduction}
The motivation to study BHs in asymptotically AdS
spacetimes is driven mostly by the gauge/gravity duality.
This correspondence~\cite{Maldacena:1997re,Aharony:1999ti} provides a powerful
framework within which certain strongly coupled gauge theories can be mapped to weakly coupled quantum gravitational 
systems in one dimension higher. 
Within this holographic setup, a BH is dual to a thermal state. Perturbed BHs are dual to non-thermal
boundary gauge theories and the approach to equilibrium in the gravitational side 
is translated to understanding thermalization in the  boundary gauge theory side~\cite{Chan:1996yk,Horowitz:1999jd,Danielsson:1999fa,Birmingham:2001pj,Son:2002sd,Kovtun:2005ev,Policastro:2002se,Friess:2006kw,Michalogiorgakis:2006jc}
(see also the review~\cite{Berti:2009kk}).

Understanding the behavior of perturbed BHs in asymptotically AdS spacetimes is therefore
a central aspect of current fundamental and practical research endeavors. 
In asymptotically flat spacetimes, the response of a BH spacetime to an external perturbation
can be roughly divided in three stages. The initial stage corresponds to on-light-cone propagation
and is entirely related to the initial data content; at intermediate times, after the initial pulse reaches the null circular geodesic,
ringdown ensues. In this stage the perturbation decays as a series of exponentially damped sinusoids, or quasinormal modes (QNMs)
whose properties are intrinsic to the BH and do not depend on the external perturbation~\cite{Berti:2009kk}; Finally, at late-times backscattering off spacetime curvature becomes important, and a power-law tail becomes visible~
~\cite{Price:1971fb,Ching:1995tj}.

It turns out that
asymptotically AdS spacetimes introduce novel aspects in the physics of BHs, mostly due to the 
absence of dissipation at infinity: the boundary is timelike and perturbations ``bounce off''.
Thus, one expects the structure of the perturbation spectrum to change completely and to even find
linear and nonlinear instabilities in these systems~\cite{Cardoso:2004hs,Uchikata:2009zz,Cardoso:2006wa,Cardoso:2013pza,Bizon:2011gg}.
A further interesting aspect of the different boundary conditions is that, for small AdS BHs (i.e., for BHs whose areal radius at the horizon is much smaller than the AdS scale) one expects imprints of asymptotically
flat spacetimes in the early signal~\cite{Barausse:2014tra,Barausse:2014pra}. This result, intuitive from a time evolution picture, seems to contradict frequency-domain analysis: the mode structure of small AdS BHs is completely different from that of say, asymptotically flat Schwarzschild or Kerr 
BHs. Furthermore, understanding the {\it linear} regime of perturbed BHs is important as benchmarks for nonlinear evolutions,
some of which have been recently reported~\cite{Bantilan:2012vu,Bantilan:2014sra,Cardoso:2014uka}. Surprisingly, time-domain evolutions in the background of rotating
BHs in AdS spacetimes are missing. This work intends to close this gap.

\section{Framework and numerical procedure}

\subsection{Wave Equation}
The massless Klein-Gordon equation in Kerr-AdS background describes scalar field perturbations in the space-time of
Kerr-AdS BHs~\cite{Cardoso:2004hs,Uchikata:2009zz}. In Boyer-Lindquist coordinates, this equation takes the form
\footnote{Notice the minor typo in Eq.(2.6) of Ref.\cite{Uchikata:2009zz}, the ${\sin^2\theta}$ in the denominator
of the second term in the r.h.s should be a ${\sin\theta}$.}
\begin{eqnarray}
\label{teuk0}
&&
-\left[{(r^2 + a^2)^2 }-\frac{\Delta_r}{\Delta_\theta}a^2\sin^2\theta\right]
         \partial_{tt}\Psi \nonumber \\
&&
+{2a\Sigma\left[({r^2}+{a^2})-\frac{\Delta_r}{\Delta_\theta}\right]}
         \partial_{t\phi}\Psi \nonumber \\
&&
+\,\Delta_{r}\partial_r\left(\Delta_{r}\partial_r\Psi\right)
+\frac{\Delta_{r}}{\sin\theta}\partial_\theta
\left({\Delta_\theta}\sin\theta\partial_\theta\Psi\right) \nonumber\\
&& 
+{\Sigma^2}\left[\frac{\Delta_{r}}{\Delta_{\theta}\sin^2\theta}-{a^2}\right] 
\partial_{\phi\phi}\Psi   = 0 ,
\end{eqnarray}
where $M$ is the BH mass, $a$ its angular momentum per unit mass, $\Delta_{r} = (r^2 + a^2)
(1 + \frac{r^2}{\ell^2}) - 2 M r$,  $\Delta_{\theta} = 1-\frac{r^2}{\ell^2}\cos^2\theta$ and $\Sigma = 1- \frac{r^2}{\ell^2}$.   
We say that a BH is small when $M/\ell\ll 1$. This definition is equivalent to requiring that the areal radius at the horizon
is small, when compared to the AdS radius~\cite{Cardoso:2013pza}.

The equation above is a linear, hyperbolic, homogeneous (3+1)D partial differential equation that was shown to yield a
well-posed initial-value problem in Refs.~\cite{Holzegel:2011qj,Andras:2009}.

\subsection{Evolution Code}
A stable evolution scheme for the Teukolsky master equation of Kerr spacetime was developed in Ref.~\cite{Krivan:1997} 
using the well-established, 2-step Lax-Wendroff numerical evolution scheme. Our evolution code uses the exact same approach, 
therefore the contents of this section are largely a review of the work presented in the literature. 

Our code uses the tortoise coordinate $r^*$ in the radial direction, which is related to the Boyer-Lindquist 
coordinates $r$ through the transformation
\begin{eqnarray}
dr^* &=& \frac{r^2+a^2}{\Delta_r}dr \; .
\end{eqnarray}
These coordinates are better suited for performing numerical evolutions in a Kerr space-time background for a number 
of reasons that are detailed in Ref.~\cite{Krivan:1997}. Next, we factor out the azimuthal dependence and use the ansatz, 
\begin{eqnarray}
\label{eq:psiphi}
\Psi(t,r^*,\theta,{\phi}) &=& e^{im{\phi}} \Phi(t,r^*,\theta) 
\end{eqnarray}
that allows us to reduce the dimensionality of the original (3+1)D Eqn.~\ref{teuk0} to a system of (2+1)D equations. 
Defining
\begin{eqnarray}
\Pi &\equiv& \partial_t{\Phi} + b \, \partial_{r^*}\Phi \; , \\
b & \equiv &
-\frac { {r}^{2}+{a}^{2}}
      { \tilde\Sigma} \; , 
\end{eqnarray}
and
\begin{eqnarray}
\tilde\Sigma^2 &\equiv & {(r^2 + a^2)^2 }-\frac{\Delta_r}{\Delta_\theta}a^2\sin^2\theta
\; 
\label{pi_eq}
\end{eqnarray} 
allows the Teukolsky equation to be rewritten in first order form as
\begin{eqnarray}
\label{eq:evln}
\partial_t \mbox{\boldmath{$u$}} + \mbox{\boldmath{$M$}} \partial_{r*}\mbox{\boldmath{$u$}} 
+ \mbox{\boldmath{$Lu$}} + \mbox{\boldmath{$Au$}} =  0 ,
\end{eqnarray}
where 
\begin{equation}
\mbox{\boldmath{$u$}}\equiv\{\Phi_R,\Phi_I,\Pi_R,\Pi_I\}
\end{equation}
is the solution vector. The subscripts $R$ and $I$ refer to the real
and imaginary parts respectively (recall that the Teukolsky function
$\Psi$ is a complex valued quantity). Explicit forms for the matrices {\boldmath{$M$}},
{\boldmath{$A$}} and {\boldmath{$L$}} can be easily computed given the expressions we 
enlist above. Rewriting Eq.\ (\ref{eq:evln}) as 
\begin{equation}
\partial_t \mbox{\boldmath{$u$}} + \mbox{\boldmath{$D$}}
\partial_{r^*} \mbox{\boldmath{$u$}}
=\mbox{\boldmath{$S$}}\; , 
\label{new_teu2}
\end{equation}
where
\begin{equation}
 \mbox{\boldmath{$D$}} \equiv \left(\begin{matrix}
                    b &   0   &  0  &  0 \cr
                    0  &   b   &  0  &  0 \cr
                    0  &   0   &  -b  &  0 \cr
                    0  &   0   &  0  &  -b \cr
                \end{matrix}\right),
\label{d_matrix}
\end{equation}
\begin{equation}
\mbox{\boldmath{$S$}} = -(\mbox{\boldmath{$M$}} - \mbox{\boldmath{$D$}})
\partial_{r^*}\mbox{\boldmath{$u$}}
- \mbox{\boldmath{$L$}}\mbox{\boldmath{$u$}} 
- \mbox{\boldmath{$A$}}\mbox{\boldmath{$u$}},
\end{equation}
and using the Lax-Wendroff iterative scheme, we obtain stable evolutions.
Each iteration consists of two steps: In the first step, the solution vector 
between grid points is obtained from
\begin{eqnarray}
\label{lw1}
\mbox{\boldmath{$u$}}^{n+1/2}_{i+1/2} &=& 
\frac{1}{2} \left( \mbox{\boldmath{$u$}}^{n}_{i+1}
                  +\mbox{\boldmath{$u$}}^{n}_{i}\right)
- \\
&  &\frac{\delta t}{2}\,\left[\frac{1}{\delta r^*} \mbox{\boldmath{$D$}}^{n}_{i+1/2}
  \left(\mbox{\boldmath{$u$}}^{n}_{i+1}
                  -\mbox{\boldmath{$u$}}^{n}_{i}\right)
- \mbox{\boldmath{$S$}}^{n}_{i+1/2} \right] \; .\nonumber
\end{eqnarray}
This is used to compute the solution vector at the next time step,
\begin{equation}
\mbox{\boldmath{$u$}}^{n+1}_{i} = 
\mbox{\boldmath{$u$}}^{n}_{i}
- \delta t\, \left[\frac{1}{\delta r^*} \mbox{\boldmath{$D$}}^{n+1/2}_{i}
  \left(\mbox{\boldmath{$u$}}^{n+1/2}_{i+1/2}
                  -\mbox{\boldmath{$u$}}^{n+1/2}_{i-1/2}\right)
- \mbox{\boldmath{$S$}}^{n+1/2}_{i} \right] \, .
\label{lw2}
\end{equation}
The angular subscripts are dropped in the above equation for clarity. All angular
derivatives are computed using second-order, centered finite difference expressions. 

We set $\Phi$ and $\Pi$ to zero on the outer radial boundary, and ``ingoing'' on the 
inner boundary (which is placed extremely close to the hole's horizon i.e. at a large negative 
value of the radial $r^*$ coordinate). Symmetries of the spherical harmonics are used to determine 
the angular boundary conditions: For even $|m|$ modes, we have $\partial_\theta\Phi =0$ at 
$\theta = 0,\pi$ while $\Phi =0$ at $\theta = 0,\pi$ for modes of odd $|m|$.
\subsection{Initial data}
We specify initial data for the evolution code by simply setting $\Phi$ to zero everywhere 
and choosing an ordinary Gaussian radial-profile for $\Pi$ and a pure spherical harmonic 
in the angular dimension. More specifically, we choose the initial data to be of the form:
\begin{equation}
\Pi \; e^{im\phi}  \; = \; e^{-{{r^*}^2}\over{2\lambda}} Y_{l m} 
\end{equation}
so that the Gaussian is centered at ${r^*} = 0$ and has a width $\lambda = 4$. We always 
choose $l=m$ in the current work.
\subsection{Parameters}
For the numerical evolutions, we use a radial grid that is uniform in the tortoise coordinate $r^*$. The inner 
boundary is placed at a large negative value of $r^*$ in a manner that it is outside of the domain for causal 
influence for the location and evolution duration of interest. The outer boundary is located precisely at 
$r^{*} = \frac{\pi}{2} \ell$, or $r=\infty$ in Boyer-Lindquist coordinates. Everything 
is computed in units of BH mass $M$ which is numerically chosen to be $1.0$. Our numerical computations 
cover a range of parameters: $a/M$ in the range of $0.0 - 0.995$; $\ell$ in the range of $20 - 10^4$; and $m$ with 
values $0$, $1$ and $2$. For simplicity we will focus on results $\ell=64, 10^4$ and $a/M=0,0.995$.
The numerical code is fully parallelized to operate on a Beowulf cluster using domain-decomposition along the 
radial dimension, and exhibits excellent scaling. We typically use radial grid resolutions of $M/40$ and angular 
resolutions of $\pi/64$, with a Courant factor of $0.5$ for the time-stepping. Each simulation was completed using 
$500$ processor-cores each, taking a few days of walltime.   
\section{Results}
%
\begin{figure}[hbtp!]
\includegraphics[width=8cm]{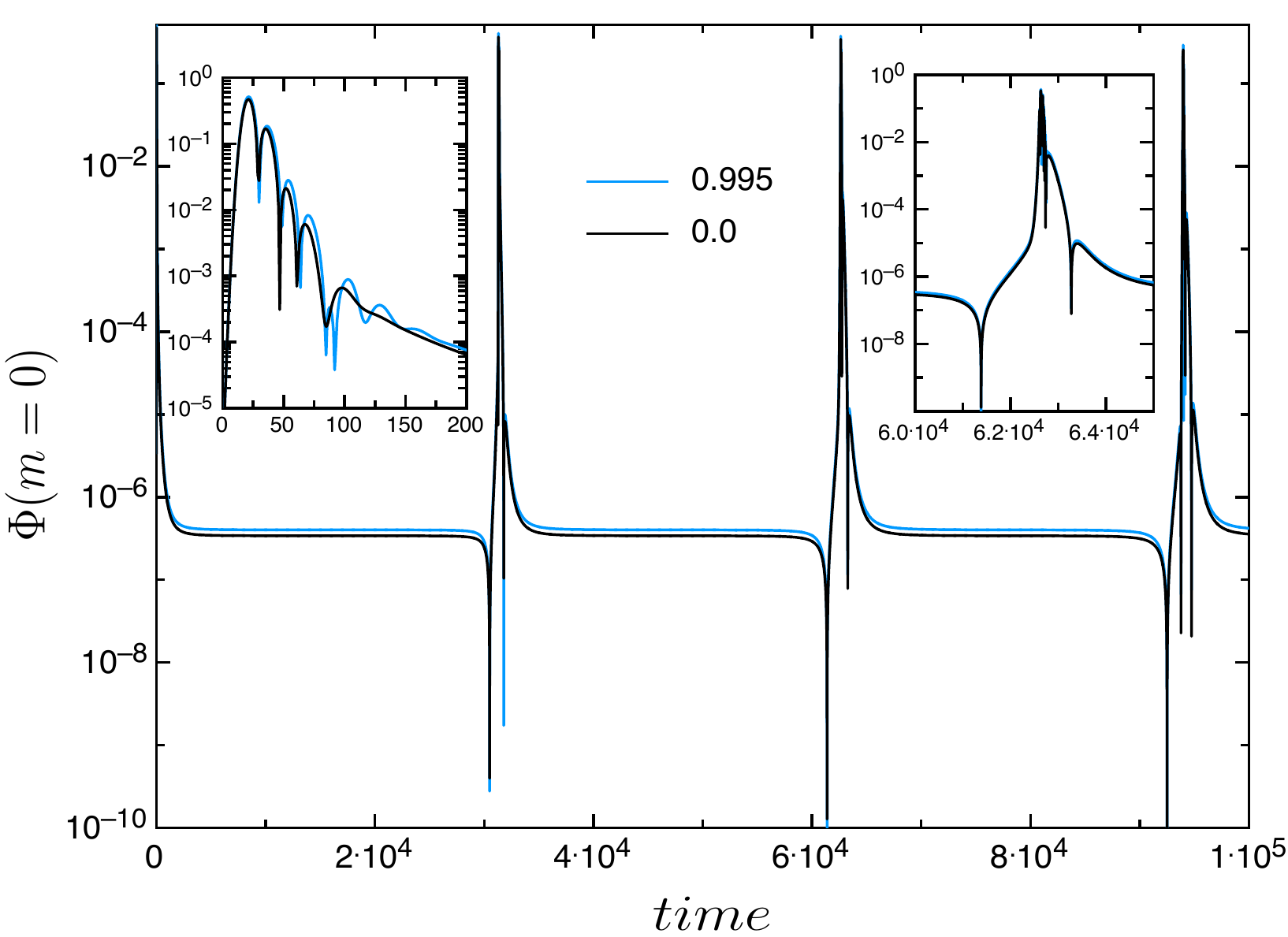}
\caption{Very long time evolution of a spherically symmetric ($l=m=0$) wavepacket in a Kerr-AdS background, with $\ell=10^{4}M$, contrasted against the evolution of the same initial wavepacket in Schwarzschild-AdS.}
\label{fig:L10000_long}
\end{figure}
%
%
\begin{figure}[hbtp!]
\includegraphics[width=8cm]{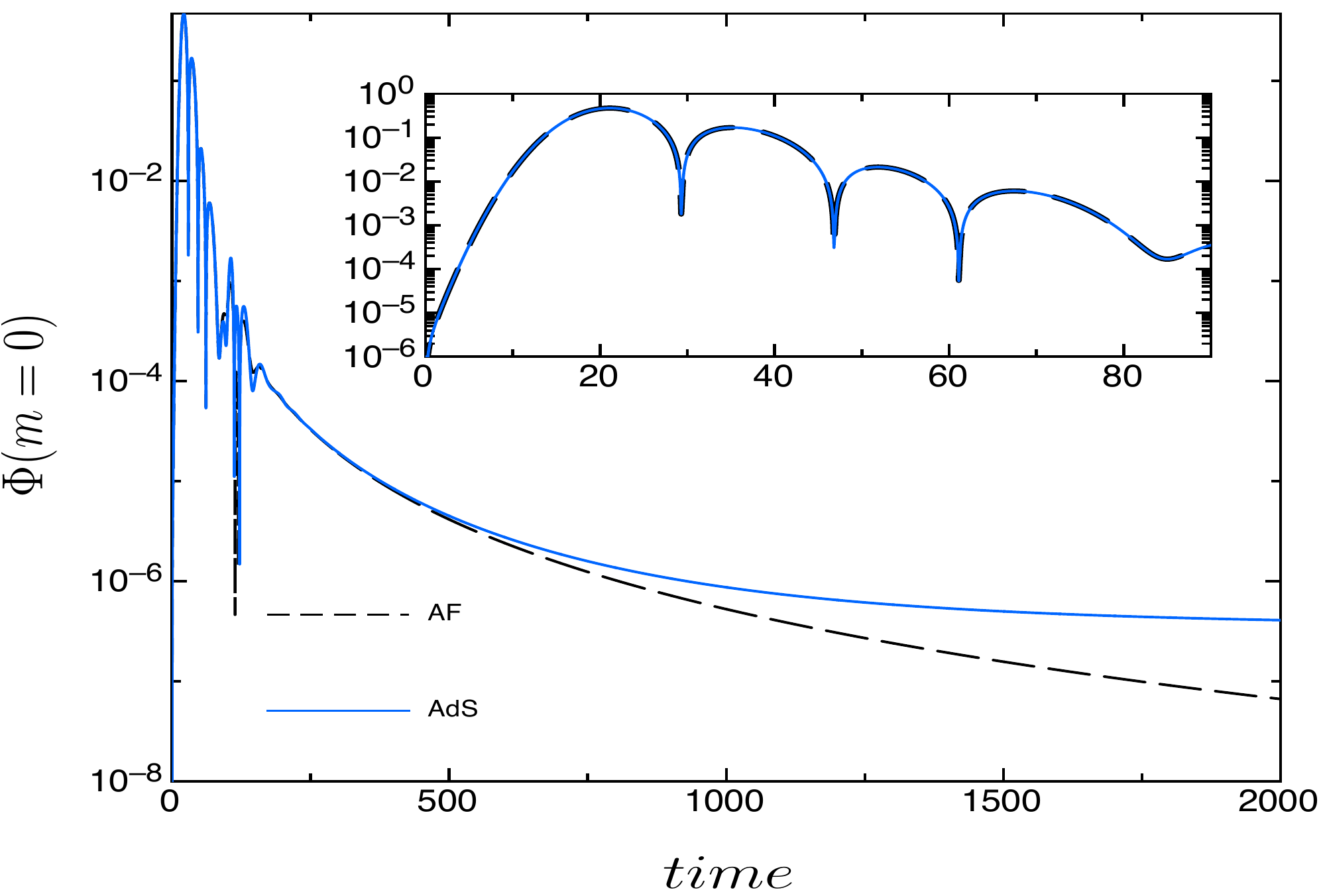}
\caption{Time evolution of a spherically symmetric ($l=m=0$) wavepacket in a small Schwarzschild-AdS background, with $\ell=10^{4}M$, contrasted against the evolution
of the same initial wavepacket in asymptotically flat spacetime. The 
pulse shows signs of Schwarzschild ringdown {\it and} power-law tail at very early times. The complete agreement with that of asymptotically BHs is shown in the inset,
which zooms-in the evolution at early times.
At very late times, not shown here, the two evolutions differ appreciably.}
\label{fig:L10000}
\end{figure}
%
%
\begin{figure}[hbtp!]
\includegraphics[width=8cm]{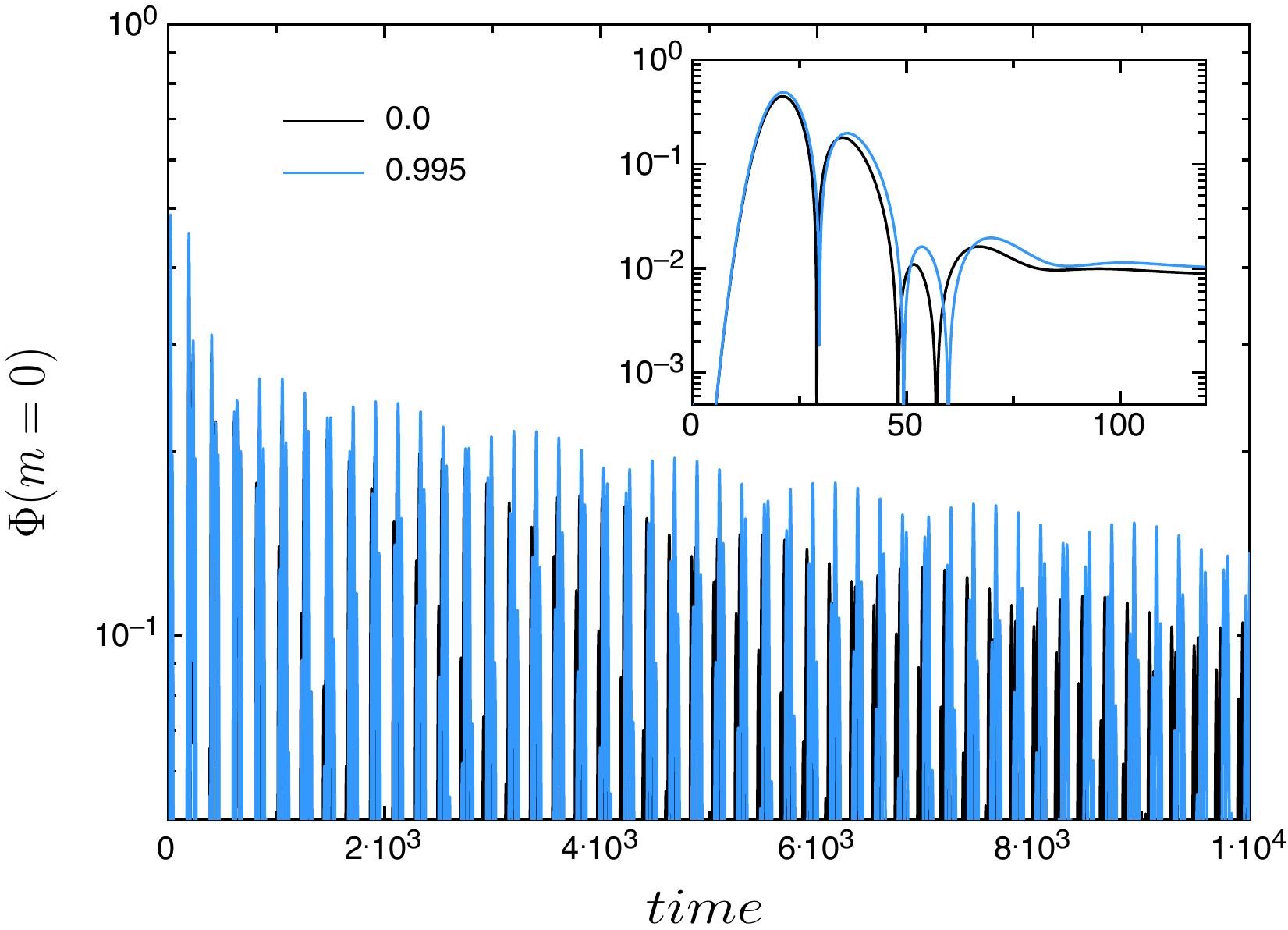}
\caption{Time evolution of a wavepacket in a small Kerr-AdS background, with $\ell=64M$,
for an initial $l=m=0$ scalar wavepacket. The BH is spinning at a rate $a/M=0$ (black line) and $a/M=0.995$ (blue line). The $m=0$
pulse ``thermalizes'' very quickly, and at late times, we find good agreement with the Kerr-AdS QNMs computed via frequency-domain 
methods~\cite{Berti:2009kk}.
The inset shows an early-time zoom-in of the evolution, which shows signs of ringdown similar to that of asymptotically flat BHs; however, before ringdown
``dies off,'' the signal has had time to travel to and back from the boundary, significantly distorting the waveform.}
\label{fig:L64m0}
\end{figure}
%
We focus now on fields extracted at the equator $\theta=\pi/2$ and at $r=20M$. A typical waveform for axially symmetric ($m=0$) fluctuations is shown in Figs.~\ref{fig:L10000_long}-\ref{fig:L64m0} for $\ell=10^4M,\,64M$. The early time evolution is dictated by the initial data and by the scattering off the BH horizon. Accordingly, the three stages -- prompt response followed by ringdown and a power-law tail -- of linearized perturbations discussed in the introduction are seen here very clearly. In fact, at early times $t \ll \ell$ and for very small $M/\ell$, the {\it quantitative details} of these stages are identical to the ones in asymptotically flat spacetime. In particular small BHs in AdS excite the same ringdown modes and tails as those in Kerr backgrounds. This is clearly seen for small BHs such as the ones in Figs.~\ref{fig:L10000_long}-\ref{fig:L10000}. A quantitative comparison of the signal with that of asymptotically flat spacetimes is provided in Fig.~\ref{fig:L10000}, where we compare the signal against that of a BH with the same mass but living in asymptotically flat spacetime. The two signals are identical up to timescales of the order of the AdS radius. Such results are not surprising: the QNMs of asymptotically flat BHs are modes leaking from the unstable null circular geodesic~\cite{Cardoso:2008bp} and therefore could be expected to be excited as well.
Perhaps the unexpected feature is that they play no role whatsoever in the spectral content of the wave operator in AdS, or in other
words, the QNMs of asymptotically flat BHs do not show up as poles of the retarded Green's function in AdS (see also Ref.~\cite{Barausse:2014tra,Barausse:2014pra}
for a related phenomena in connection with BHs in astrophysical environments). As is well known,
a frequency-domain analysis for the QNMs of small BHs in AdS returns extremely long-lived modes lying close to the purely
real, AdS normal modes 
\begin{equation}
\omega \ell=3+l+2n\,,\label{AdS:freq}
\end{equation}
with $l,n=0,1,2,..$ angular and radial numbers respectively~\cite{Berti:2009kk,Cardoso:2013pza}.

To summarize, imprints of asymptotic flatness are present in waveforms of asymptotically AdS BHs. The overall magnitude of these imprints can be varied, by appropriately tuning the initial data (for example, it can be chosen in such a way that ringdown excitation is suppressed (see Refs.~\cite{Andersson:1995zk,Berti:2006wq} for studies in frequency-domain and Refs.~\cite{Andersson:1999wj,Glampedakis:2001js,Andersson:1998swa,Bernuzzi:2008rq} for corresponding analysis in the time-domain), but is otherwise generic. This is one of the main messages of our work.

Because the AdS radius introduces an extra length scale in the problem, the evolution is characterized by two extra stages relatively to its asymptotically flat counterpart. The first occurs on timescales
large enough that AdS curvature becomes important, modifying the characteristic of the ``tail''. Because the power-law tails
are a large wavelength effect, they feel the impact of the AdS modifications and our results suggest that these changes will eventually
contribute to the settling in of the AdS resonant modes. The final stage is the reflection off the boundary at $t\sim \ell\pi/2$.
These two stages are apparent already in Fig.~\ref{fig:L10000_long}, where the reflection off the boundary is visible.
However, a smooth transition to the AdS regime would take a large number of reflections and consequently long timescales. We therefore focus on a somewhat larger AdS BH, with $\ell=64M$, shown in Fig.~\ref{fig:L64m0} for $a=0,\,0.995M$. 
Furthermore, the inset in Fig.~\ref{fig:L10000} shows that ringdown lasts for $\sim 80M$; thus, similar initial data will result in a ringdown stage
similar to that of asymptotically flat BHs, as long as $l\gg 80M$, i.e., as long as the signal does not have time to travel back from the boundary.
Thus, we expect that with $\ell=64M$ the ringdown stage will not be identical to its asymptotically flat counterpart.

Both these features are borne out of our results, shown in Fig.~\ref{fig:L64m0}. 
The inset, zooming in on the evolution at early times, shows that the ``asymptotically-flat'' memory (i.e., ringdown characteristic of asymptotically flat BHs)
is quickly distorted; the reason, as we explained, is that on timescales of $\sim80M$ the field bounced back from the boundary already.
We find that the transition to the AdS regime occurs after a few reflections off the boundary. 
Notice that upon each reflection at the boundary the wave is partially absorbed by the BH, leading to a steady decrease in amplitude~\footnote{In fact, this simple picture of waves at the boundary and consequent absorption at the horizon can be used to estimate the decay rate of any fluctuation in a confined geometry solely from the knowledge of the absorption probabilities of {\it asymptotically flat} BHs~\cite{Brito:2014nja}}. We find that both the resonant frequencies and damping times are well approximated by frequency-domain results~\cite{Berti:2009kk}; in particular, for small BHs one recovers~\eqref{AdS:freq}. For example, a Fourier analysis
of the $m=0, a=0$ waveform yields $\ell \omega\sim 2.98-i0.0045$, in very good agreement with numerical and analytical results in the frequency-domain, which yield $\ell \omega\sim2.92-i0.0049$~\cite{Cardoso:2004hs,Uchikata:2009zz}. There are no power-law tails appearing at time scales
much larger than $\ell$ because of the decay properties of the potential~\cite{Ching:1995tj,Horowitz:1999jd}.

%
\begin{figure*}[hbtp!]
\includegraphics[width=8cm]{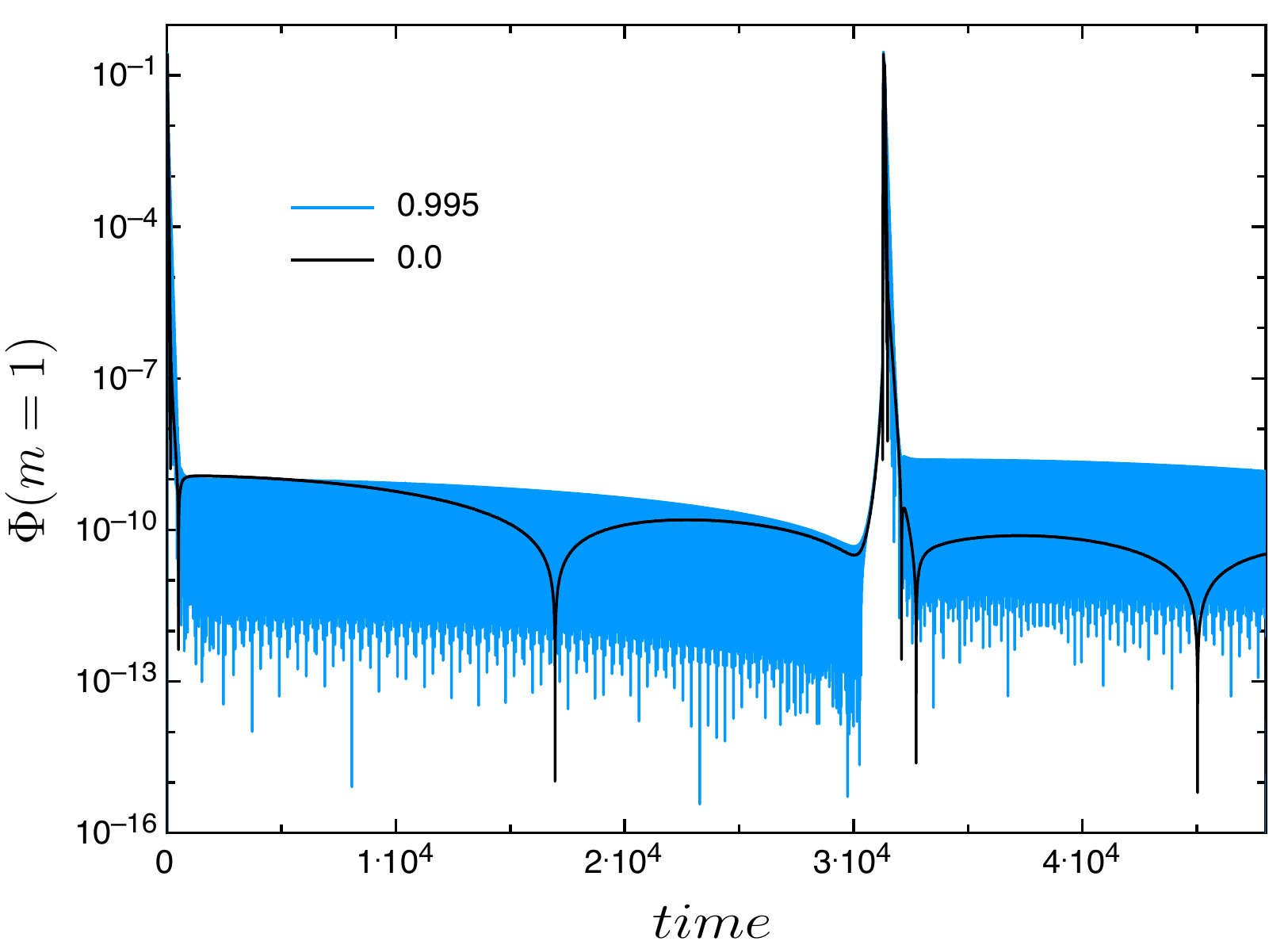}\hskip 1cm
\includegraphics[width=8cm]{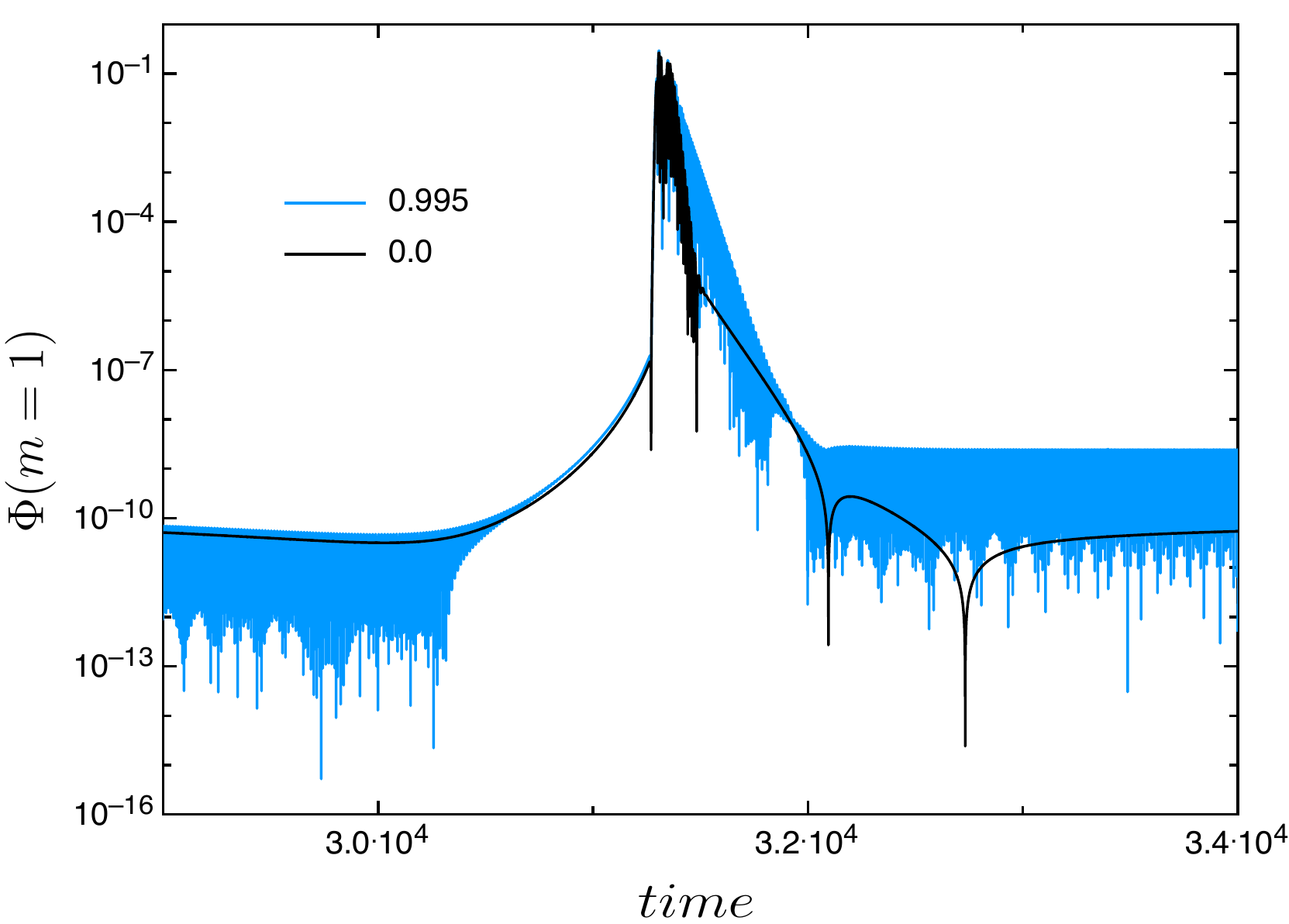}
\caption{Time evolution of a dipolar $l=1$ wavepacket in a small Kerr-AdS background, with $\ell=10^{4}M$. 
Left panel shows $m=1$ waveforms in the background of a BH with spin $a=0$ and $a=0.995M$. The right panel zooms in on the corresponding signal at late times.
The waveform begins with a high frequency part corresponding to a {\it Schwarzschild} ringdown. At timescales $t\sim \ell$ the effects of the cosmological constant are felt and a low frequency component eventually settles in before complete reflection at the boundary. Eventually, the low frequency component dominates the signal. The turn-over to the AdS, with ``thermalization'' at low frequencies is more clearly seen in Fig.~\ref{fig:L64m0}.}
\label{fig:L1000m}
\end{figure*}
\begin{figure}[hbtp!]
\includegraphics[width=8cm]{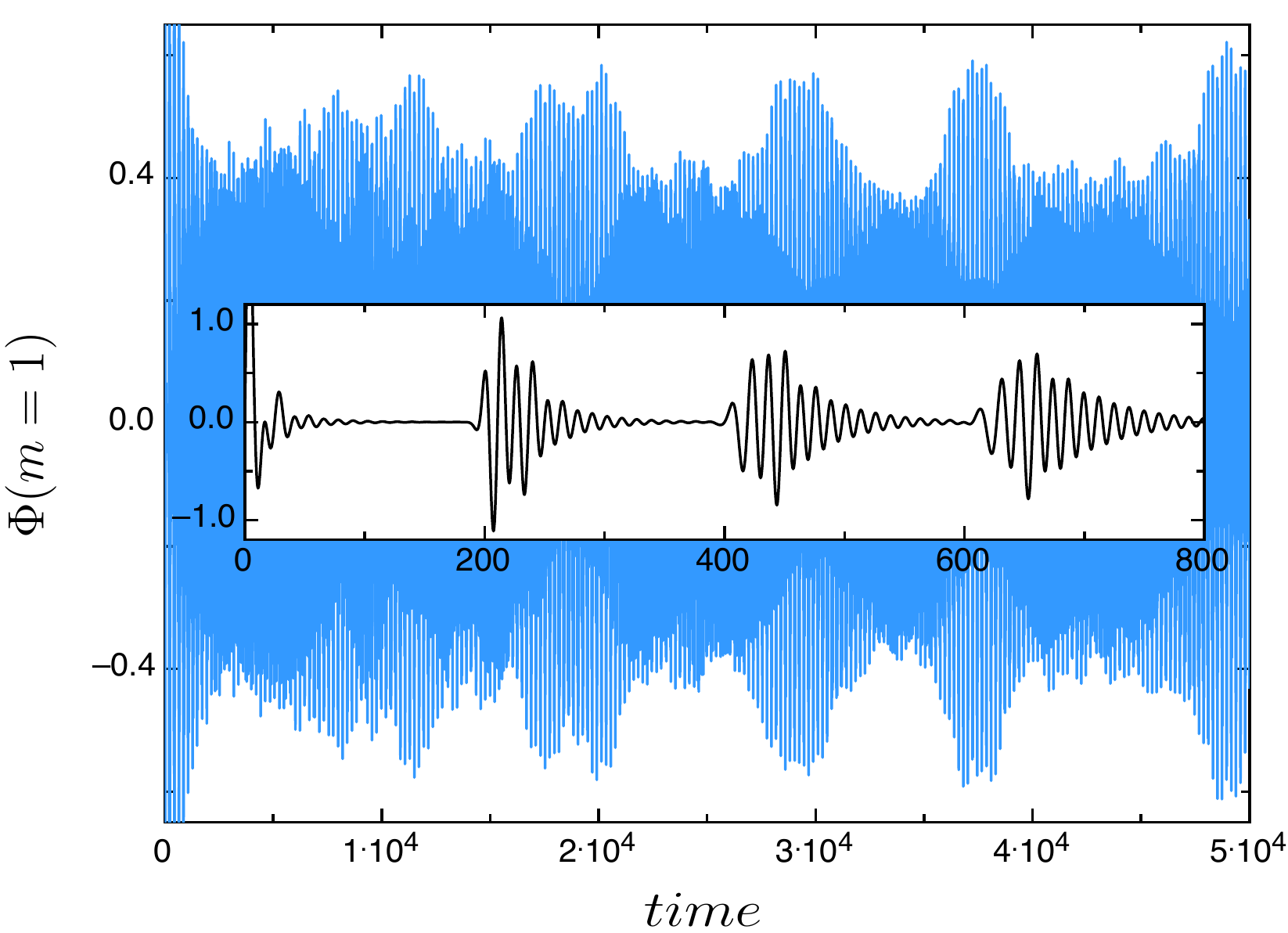}
\caption{Very long time evolution of an $m=1$ wavepacket in a Kerr-AdS background, with $\ell=64M$ and $a/M = 0.995$, contrasted against the evolution of the same, however, with $m=0$. Instability on the scale of ${10^6}M$ develops in the non-axisymmetric case, as predicted by frequency-domain analysis~\cite{Cardoso:2004hs,Uchikata:2009zz,Cardoso:2006wa,Cardoso:2013pza}. }
\label{fig:instability}
\end{figure}
The overall picture is the same also for non-axisymmetric modes, whose evolution is summarized in Figs.~\ref{fig:L1000m}-\ref{fig:instability}.
However, because these modes are non-axisymmetric, they are prone to superradiant effects when interacting with the BH~\cite{Cardoso:2004hs,Uchikata:2009zz,Cardoso:2006wa,Cardoso:2013pza,Brito:2014nja}: for small enough rotating BHs in AdS, the resonant frequency
is that of pure AdS, $\omega \sim 1/\ell$ and can be superradiant, i.e.,
\begin{equation}
\omega <m\Omega\,,
\end{equation}
with $\Omega$ the horizon's ``angular velocity.'' In this regime, after reflection at the boundary the wave is {\it amplified} close to the horizon, leading to an exponential growth of the field. This is usually known as a BH bomb or superradiant instability~\cite{Cardoso:2004hs,Uchikata:2009zz,Cardoso:2006wa,Cardoso:2013pza,Brito:2014nja,Shlapentokh-Rothman:2013ysa}. Figure~\ref{fig:instability} shows indication that the evolution of wavepackets around small, rapidly rotating BHs in AdS does lead to a growth of the field. The initial stages show the same features as thhose already discussed, with the initial pulse scattering off the BH and bouncing of the AdS boundary. This process will typically filter high-frequency components off the signal, because they are easily absorbed by the BH. At intermediate-late stages, the signal consists mainly on lower-frequency components which are prone to superradiance growth. We estimate the growth-timescale to be of order $4\times 10^6M$, which agrees with the estimate of Ref.~\cite{Brito:2014nja} and with frequency-domain calculations~\cite{Cardoso:2004hs,Uchikata:2009zz,Cardoso:2006wa,Cardoso:2013pza}.

\section{Conclusions}

The early-time evolution of fluctuations around small AdS BHs is {\it identical} to that of their asymptotically flat counterpart:
the evolution can be divided in the same three stages, prompt response, ringdown and late-time power-law tail. 
We highlight that these stages are {\it not} captured by a frequency-domain analysis~\cite{Barausse:2014tra,Barausse:2014pra}. What frequency-domain
analysis does capture are the late-time dynamics, when the fluctuations have had time to bounce off the boundary several times and eventually decay in the Kerr-AdS quasinormal modes. These modes decay in time and can be understood as a sequence of reflections and subsequent absorption at the BH horizon. Finally, we have have indications -- directly in the time-domain -- that small, rotating BHs in AdS are unstable against a superradiant mechanism in the sense that small non-axisymmetric fluctuations grow~\cite{Cardoso:2004hs,Uchikata:2009zz,Cardoso:2006wa,Cardoso:2013pza,Shlapentokh-Rothman:2013ysa}.

The dynamics in asymptotic AdS spacetimes are rich but complex. Their understanding requires the full nonlinear machinery of Einstein equations,
complemented with robust linearized analysis. Our study intends to partially fill the gap in linearized, time-evolution studies in asymptotic AdS
spacetimes.

{\bf \em Acknowledgements.}
%
V.C. acknowledges financial support provided under the European Union's FP7 ERC Starting Grant ``The dynamics of black holes:
testing the limits of Einstein's theory'' grant agreement no. DyBHo--256667.
This research was supported in part by Perimeter Institute for Theoretical Physics. 
Research at Perimeter Institute is supported by the Government of Canada through 
Industry Canada and by the Province of Ontario through the Ministry of Economic Development 
$\&$ Innovation.
This work was supported by the NRHEP 295189 FP7-PEOPLE-2011-IRSES Grant.
G.K. acknowledges research support from NSF Grants No. PHY-1016906, No. CNS-0959382, No. PHY-1135664, 
and No. PHY-1303724, and No. PHY-1414440, and from the U.S. Air Force Grant No. FA9550-10-1-0354 and No. 10-RI-CRADA-09. Some 
of the long duration computations were performed on the UMass shared cluster of the MGHPCC green 
high-performance computing facility. 
%


\end{document}